\documentclass[reprint,aps,amsmath,amssymb,twoside,prd,showkeys,superscriptaddress]{revtex4-1}
\pdfoutput=1
\usepackage{verbatim}
\usepackage{comment}
\usepackage{graphicx}
\usepackage[T1]{fontenc}
\usepackage{graphicx}
\usepackage{epsfig}
\usepackage{bm}
\usepackage{tensor}
\usepackage{amsfonts}
\usepackage{amssymb}
\usepackage{float}
\usepackage{amsmath}
\usepackage{subfigure}
\usepackage{dcolumn}
\usepackage{cancel}
\usepackage[colorlinks]{hyperref}
\usepackage[utf8]{inputenc}

\usepackage[usenames,dvipsnames]{color}
\hypersetup{
     breaklinks=true,
    pdfstartview={FitH},    % fits the width of the page to the window
    colorlinks=true,       % false: boxed links; true: colored links
    linkcolor=blue,          % color of internal links
    citecolor=red,        % color of links to bibliography
    filecolor=magenta,      % color of file links
    urlcolor=blue,           % color of external links
    anchorcolor=green,      % Color for anchor text
    linktocpage=true
}

\def\doi{http://doi.org}

\newcommand{\be}{\begin{equation}}
\newcommand{\ee}{\end{equation}}

\newcommand{\HCd}{\mathcal{H}}

\def\HCdt0{\tilde{\HCd}_{0}}

\newcommand{\affcam}{DAMTP, Centre for Mathematical Sciences, University of Cambridge, Wilberforce Road, Cambridge CB3 0WA, United Kingdom}
\newcommand{\affcamast}{Kavli Institute of Cosmology (KICC), University of Cambridge, Madingley Road, Cambridge, CB3 0HA, UK}
\newcommand{\affmColCam}{Queens’ College, Cambridge, CB3 9ET, U.K.}

\begin{document}

\title{Dark energy interactions near the galactic centre}
\author{David Benisty}
\email{db888@cam.ac.ukl}
\affiliation{\affcam}\affiliation{\affcamast}\affiliation{\affmColCam}
\author{Anne-Christine Davis}
\email{ad107@cam.ac.uk}
\affiliation{\affcam}\affiliation{\affcamast}
\begin{abstract}
We investigate scalar-tensor theories, motivated by dark energy models, in the strong gravity regime around the black hole at the centre of our galaxy. In such theories general relativity is modified since the scalar field couples to matter. We consider the most general conformal and disformal couplings of the scalar field to matter to study the orbital behaviour of the nearby stars around the galactic star center $Sgr A^{*}$. Markov Chain Monte Carlo (MCMC) simulation yields a bound on the parameters of the couplings of the scalar field to matter. Using Bayesian Analysis yields the first constraints on such theories in the strong gravity regime.
\end{abstract}
\maketitle
\section{Introduction}
\label{sec:Introduction}

The discovery that the Universe is undergoing a period of accelerated expansion today requires a modification of General Relativity (GR). This could be the addition of a Cosmological Constant $\Lambda$, \cite{Perlmutter:1998np,Weinberg:1988cp,Lombriser:2019jia,Copeland:2006wr,Frieman:2008sn,Riess:2019cxk}. This, however,
requires severe fine tuning by and is viewed as theoretically unsatisfactory. 
Another theoretical mechanism developed is the quintessence model  \cite{Starobinsky:1979ty,Starobinsky:1980te,Guth:1980zm,Albrecht:1982wi,Mukhanov:1981xt,Guth:1982ec,Linde:1981mu,Barrow:1988xh,Barrow:1988xi,Elizalde:2008yf,Ratra:1987rm,Caldwell:1997ii,Zlatev:1998tr,Caldwell:1999ew,Chiba:1999ka,Bento:2002ps,Tsujikawa:2013fta,Caldwell:1997ii,Ratra:1987rm,Peebles:1987ek,Barreiro:1999zs,Carroll:1998zi,Chiba:1999wt,Sahni:1999qe}, where a scalar field with a slow roll behaviour is responsible for the late time acceleration of our universe. In such models the scalar field does not couple to ordinary matter, either directly or by quantum corrections.  
Other possibilities include a modification of general relativity itself. 
Such possibilities include modifying the Einstein-Hilbert action to become a function of the Ricci scalar, the so-called f(R) theories, or the addition of a scalar field which couples to matter, the scalar-tensor theories. The coupling between the scalar field and matter is derived using theoretical considerations, involving a transformation from Jordan to Einstein frames, see \cite{CANTATA:2021ktz} for a recent review.  Bekenstein has shown the most general coupling of a scalar field to matter is via a conformal and disformal transformation \cite{Bekenstein:1992pj}. Scalar-tensor theories with a conformal coupling can be rewritten in terms of $f(R)$ and vice versa \cite{ Brax:2008hh}. However, such theories give rise to fifth forces which are subject to strict limits by solar system tests of general relativity \cite{Bertotti:2003rm}. Consequently the effect of fifth forces need to be screened in the solar system, giving rise to screened modified gravity models whereby the effects of the scalar field depends on the environment. Such models can be screened via the chameleon mechanism \cite{Khoury:2003aq, Brax:2004qh},  symmetron \cite {Hinterbichler:2010es}, Vainshtein \cite{ Vainshtein:1972sx} or Damour-Polyakov mechanism \cite{ Damour:1994zq,Brax:2010gi}. All rely in different respects on the environment such that the fifth force is screened in the solar system and the theory evades all solar system tests but can give rise to modifications to GR on cosmological scales \cite{Damour:1992we,Julie:2017pkb,Bertotti:2003rm,Williams:2004qba,Khoury:2003aq,Damour:1994zq,Vainshtein:1972sx,Babichev:2009ee}.   
%The effects of a conformal coupling must be suppressed in the solar system in order to comply with gravitational tests for the existence of fifth forces or strong equivalence principle violation %The resulting bounds on the conformal coupling are severe and screening mechanisms have been invoked in order to comply with rather unnaturally low coupling 

Similarly the disformal coupling to matter gives rise to modifications to GR which can be constrained in the solar system\cite{Brax:2018bow} and in collider physics\cite{Brax:2015hma}. This results in constraints on the disformal coupling to matter. \cite{Koivisto:2008ak,Zumalacarregui:2010wj,Koivisto:2012za,vandeBruck:2013yxa,Brax:2013nsa,Neveu:2014vua,Sakstein:2014isa,Sakstein:2014aca,Desmond:2019ygn}. 

Whilst solar system physics puts constraints on any modified gravity model theories with screening can still give observational effects. For example whilst the fifth force is suppressed in the solar system they give rise to deviations on cosmological scales from the standard GR predictions and could modify the growth of structures. Such deviations could be detected in future satellite experiments \cite{Vagnozzi:2021quy,Brax:2020hkc}. 

It is also possible to constrain a class of modified gravity models, namely those that screen with the chameleon \cite{Yoo:2012ug,Karwal:2021vpk} or symmetron mechanism  \cite{Hinterbichler:2011ca}, using laboratory experiments \cite{Brax:2018iyo,Elder:2019yyp,Brax:2019rwf,Brax:2020hkc}. Indeed such experiments give tight constraints on the model parameters. This means that this class of models are constrained by observation on very different scales and different regimes. 

Whilst modified gravity models have been tested on both small and solar system scales and their effects on cosmology computed in some cases they have yet to be tested in the strong gravity regime. This is the goal of our work. It is known that there is a relativistic, compact object in the galactic centre; the black hole $Sgr A^{*}$.  The stars orbiting $Sgr A^{*}$ are called the S-stars\cite{Yu:2016nzn,2016ApJ83017B,Abuter:2018drb,2009ApJ707L114G,Do:2019txf,Abuter:2020dou,Amorim:2019hwp,2017ApJ...845...22P}. A large fraction of these stars have orbits with high eccentricities. Thus, they reach high velocities at the pericenter and can be used for constraining modified gravity
\cite{Will:2018ont,Will:1997bb,Scharre:2001hn,Moffat:2005si,Zhao:2005zq,Bailey:2006fd,Deng:2009tg,Barausse:2012da,Borka:2012tj,Enqvist:2013tsa,Borka:2013dba,Capozziello:2014rva,Berti:2015itd,Borka:2015vqa,Zakharov:2016lzv,Zhang:2017srh,Dirkes:2017ecu,Pittordis:2017byg,Hou:2017cjy,Nakamura:2018yaw,Banik:2018ydl,Dialektopoulos:2018iph,Kalita:2018ubo,Will:2018ont,Banik:2019zme,Pittordis:2019kxq,Nunes:2019bjq,Anderson:2019eay,Gainutdinov:2020bbv,Bahamonde:2020bbc,Banerjee:2020rrd,Ruggiero:2020yoq,Okcu:2021oke,deMartino:2021daj,DellaMonica:2021xcf,DAddio:2021xsu}.

%\cite{me and Phil}
Recently \cite{Brax:2018bow} have extended the programme of \cite{Damour} to include both conformal and disformal couplings to matter in scalar-tensor theories. Rather than specialising to a specific model the programme uses a general scalar-tensor theory with parameters chosen to evade known constraints and computes physical quantities using a PN expansion. This enables such theories to be tested in new regimes. Whilst previous work has concentrated on considering such theories in the solar system in this work we extend the tests to the strong gravity regime around the region of $Sgr A^{*}$.

The plan of the work is as follows: Section \ref{sec:tbp} formulates the two body equations of motion with the conformal and the disformal interaction. Section \ref{sec:method} explains the method. Section \ref{sec:Res} explains the results and section \ref{sec:Dis} discusses the results in light of model comparison. 
\section{Formalism}
\label{sec:tbp}
Many modified gravity models can be written in the schematic form:
\be
\begin{split}
\mathcal{L}=\frac{R}{2\kappa^2}-\frac{1}{2} g^{\mu\nu} \phi_{,\mu} \phi_{,\nu} -V(\phi) +\mathcal{L}_m(\psi_i, g_{\mu\nu})    
\end{split}
\ee
where the scalar field $\phi$. \cite{Bekenstein:1992pj} shows that the most general theory of a scalar coupled to matter:
\begin{equation}
\bar{g}_{\mu\nu}=A(\phi,\partial_\nu\phi)^2g_{\mu\nu}+B(\phi,\partial_\nu\phi)^2\partial_\mu\phi\partial_\nu\phi.
\end{equation}
The matter fields are denoted by $\psi_i$ and their Lagrangian is $\mathcal{L}_m$. \cite{Brax:2018bow} focuses of nearly massless scalar and takes the form $V(\phi)  = 0$ since dark energy, which is negligible in the galactic star centre system. As a simple anzats the conformal factor is studied:
\be
A(\phi)= \exp\left[\beta \phi/\kappa\right], \quad B(\phi)^2 = 2/\kappa^2\Lambda^2, 
\ee
that is characterised by the constant coupling $\beta$ and the disformal interaction is specified by the suppression scale $\Lambda$. \cite{Brax:2013uh,Brax:2018bow,Kuntz:2019plo} use the center of mass coordinates, with the reduced mass $\mu$ and the total mass of the binary system $M$. The reduced action per $\mu$ reads:
\begin{equation}
\mathcal{L} = \mathcal{L}_0 + \frac{1}{c^2} \mathcal{L} + \mathcal{L}_{Dis} + \mathcal{O}(\frac{1}{c^4}),
\end{equation}
where:
\begin{subequations}
\begin{equation}
\mathcal{L}_0 = \frac{1}{2}v^2 + \frac{G M}{r} (1+2\beta^2),
\end{equation}
\begin{equation}
\mathcal{L}_1 = \frac{v^4}{8} + \frac{G M}{2 r}\left((3-2 \beta^2)v^2 - (1+2 \beta^2)^2\frac{G M }{r}\right),
\end{equation}
\begin{equation}
\mathcal{L}_{Dis} = \frac{4 \beta^2 G M^2}{\Lambda^4 r^4 } \left(v^2 -2  (v\cdot \hat{n})^2 \right).
\end{equation}
\end{subequations}
The action is to leading order in the first Post Newtonain (PN) order. When $\beta$ and $1/\Lambda$ go to zero the action reduces to the standard Einstein-Infeld-Hoffmann (EIH) action \cite{AIHPA_1985__43_1_107_0,Blanchet:2011wga}. The center of mass coordinate system is defined by:
\begin{equation} 
\begin{split}
\vec{r} := \vec{r}_1 - \vec{r}_2, \quad \vec{v} := \dot{\vec{r}} 
\\M := M_1 + M_2, \quad \mu := \frac{M_1 M_2}{M_1 + M_2}, \quad \nu := \frac{\mu}{M} 
%\quad X := \frac{M_1 r_1 + M_2 r_2}{M}
%\quad W := \frac{M_1 \dot{r}_1 + M_2 \dot{r}_2}{M} 
\end{split}
\end{equation}
with $0 \leq \nu \leq 1/4$. Since the research focuses on the galactic star center, we set $\nu = 0$ which demonstrates that the masses of the S stars is much lower then the mass of the black hole. 

\begin{figure}[t!]
 	\centering
\includegraphics[width=0.49\textwidth]{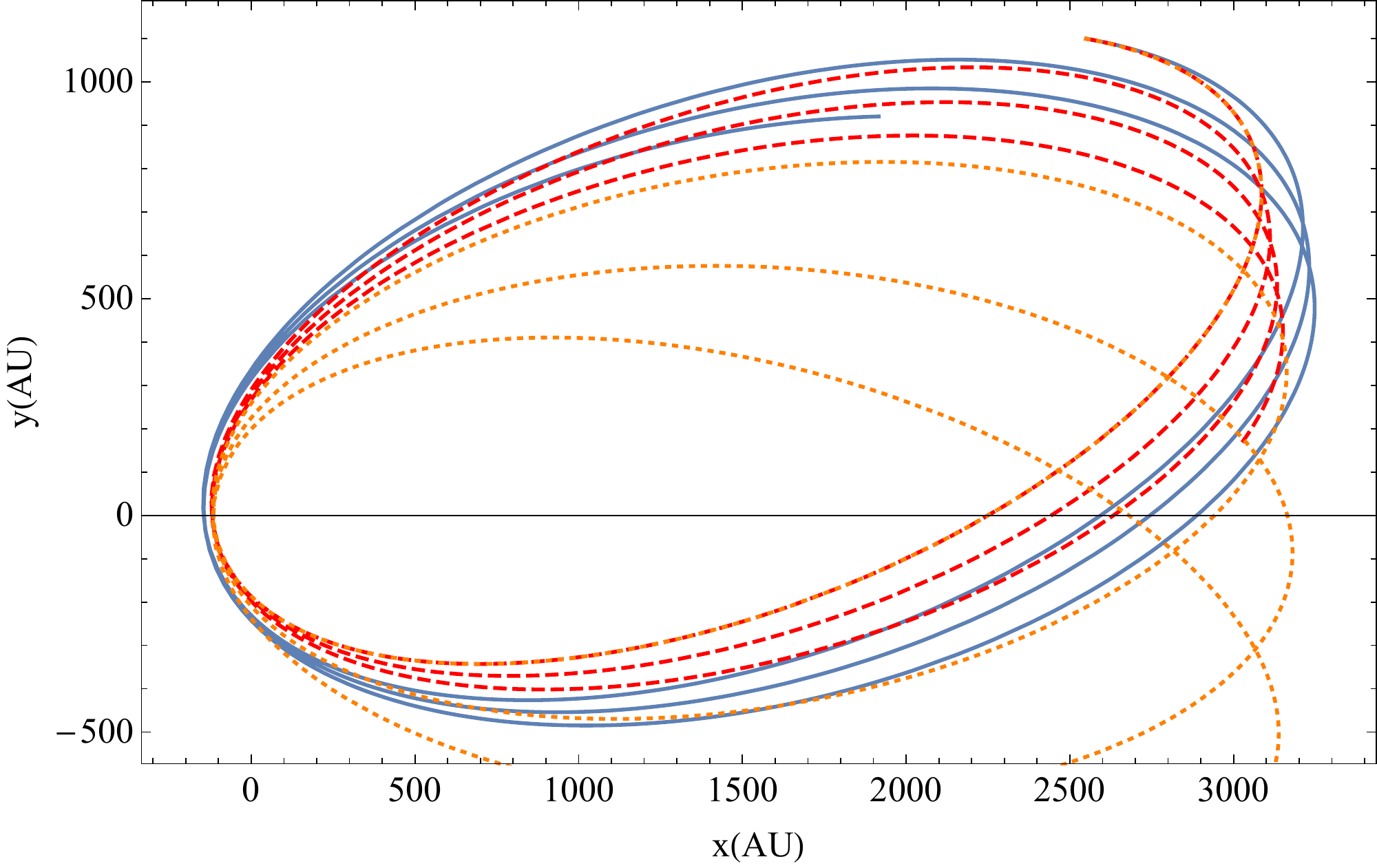}
\caption{\it{30 years simulated motion of the S2 star for the GR solution (smooth blue line), with conformal interaction where $\beta^2 = 0.1$ (dashed red line) and with the disformal interaction where $1/\Lambda^4 = 0.01\, AU^3/10^6 M_{\odot}$ (dotted orange line). }}
 	\label{fig:exam}
\end{figure}

\begin{figure}[h]
 	\centering
\includegraphics[width=0.44\textwidth]{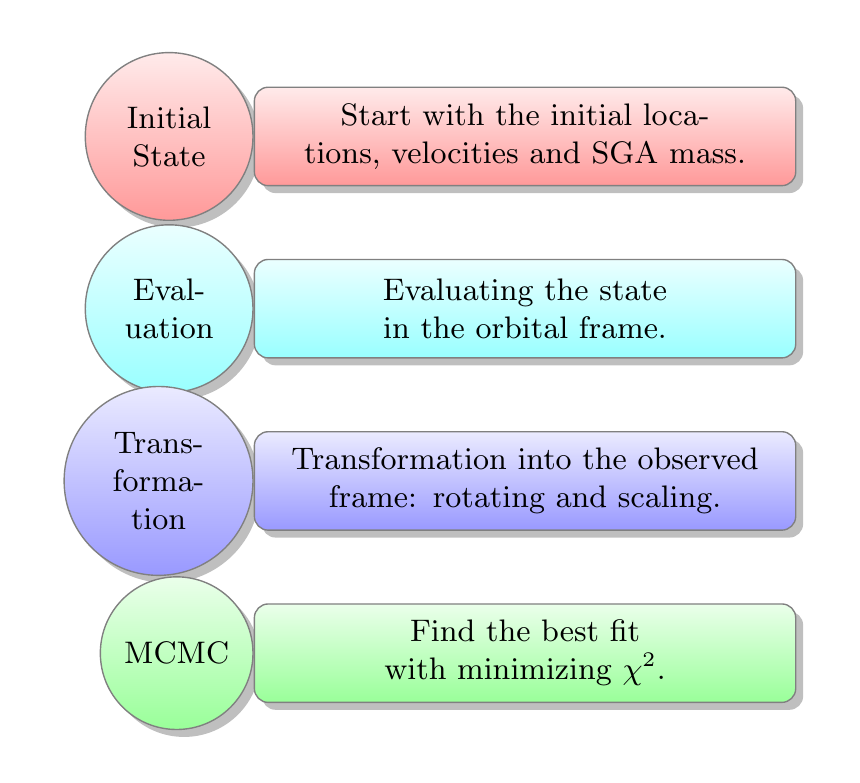}
\caption{\it{Points of the method procedure in our analysis starting from the initial conditions evaluation up to minimizing the corresponding $\chi^2$ that fits with the location and the velocities data of S2 and S38.}}
 	\label{fig:meth}
\end{figure}

The variation with respect to $\vec{r}$ yields:
\begin{equation}
\begin{split}
\ddot{x} = -\frac{G M x}{r^3}\\ + \frac{G M}{c^2 r^4} \left(x \left(4 G M+r \left(3 v_x^2-v_y^2\right)\right)+4 r v_x v_y y\right), 
\label{eq:eqMot}
\end{split}
\end{equation}
with the contributing of the conformal coupling $\beta$:
\begin{equation}
+\beta^2 \frac{2G M x}{c^2 r^4}\left(r \left(v^2-c^2\right)+2 G M\right) ,
\end{equation}
and the contribution of the disformal coupling $\Lambda$:
\begin{equation}
\begin{split}
-\frac{4 G M^2}{\Lambda^4 r^8} \left(G M r x+2 (v_x x+v_y y) \left(v_x \left(x^2-2 y^2\right)+3 v_y x y\right)\right) \\
+ \mathcal{O}(c^{-2},\beta^2,\Lambda^{-8}).   
\end{split}
\end{equation}
The same equation is satisfied for the $y$ axis, with the exchange of $x$ into $y$ and vice versa:
\begin{equation}
x \leftrightarrow y, \quad v_x \leftrightarrow v_y.
\end{equation}
Again, when $\beta$ and $1/\Lambda$ go to zero the equation of motion reduces to the standard GR solution in the first PN limit. In order to demonstrate the effect of the conformal and the disformal transformations on the S2 motion, we solve Eq. (\ref{eq:eqMot}) numerically. Fig \ref{fig:exam} shows the numerical solution of the S2 orbit in the reference frame, where the location of $Sgr A^{*}$ is $(0,0)$. The numerical solution is for 30 years. With large conformal interaction $\beta^2 = 0.1$ (dashed red line) and with large disformal interaction strengths, where $1/\Lambda^4 = 0.01$, the procession is much bigger. Therefore the possible range for these term has to be much lower as we will see from the Bayesian Analysis.
\begin{figure*}[t!]
 	\centering
\includegraphics[width=0.45\textwidth]{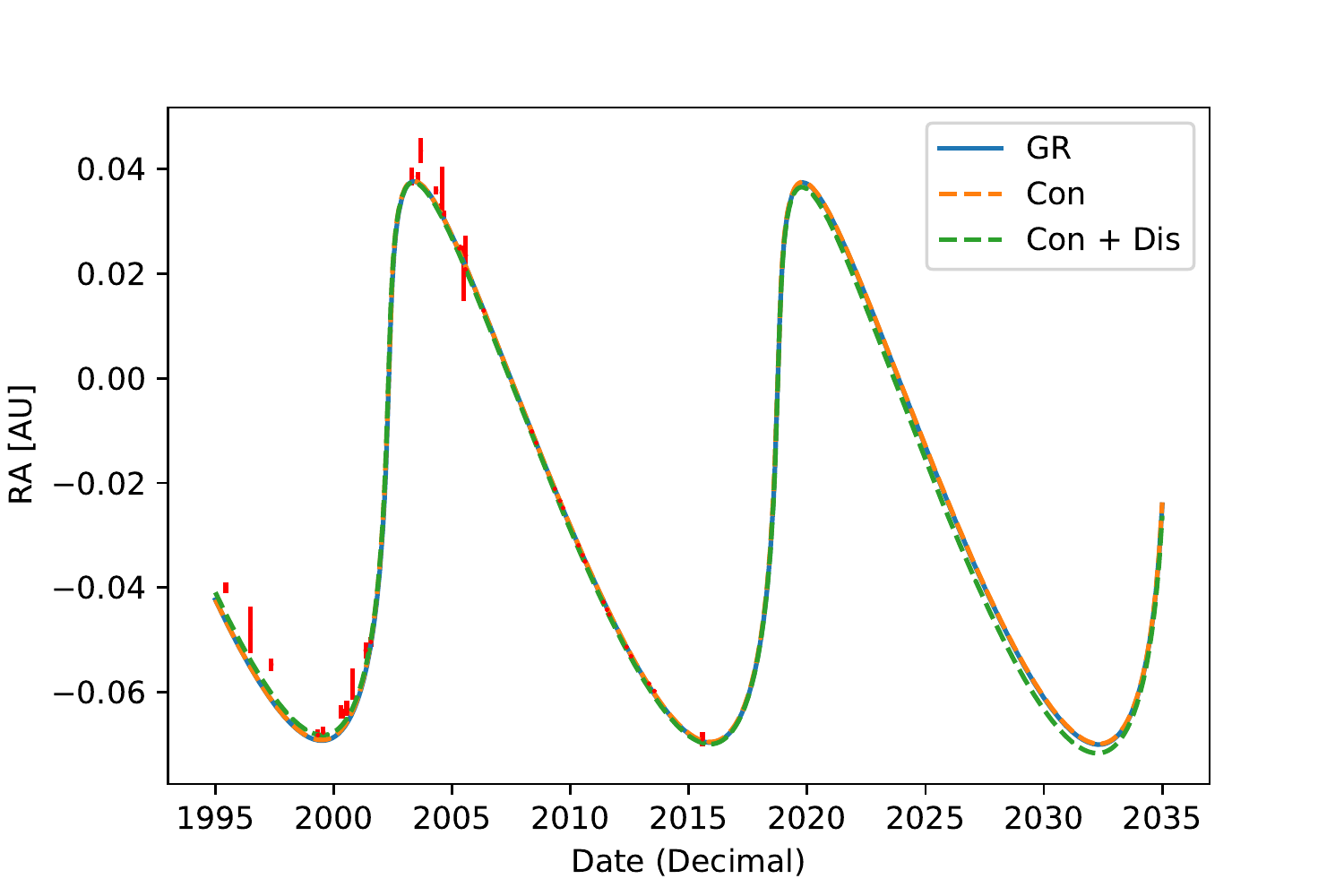}
\includegraphics[width=0.45\textwidth]{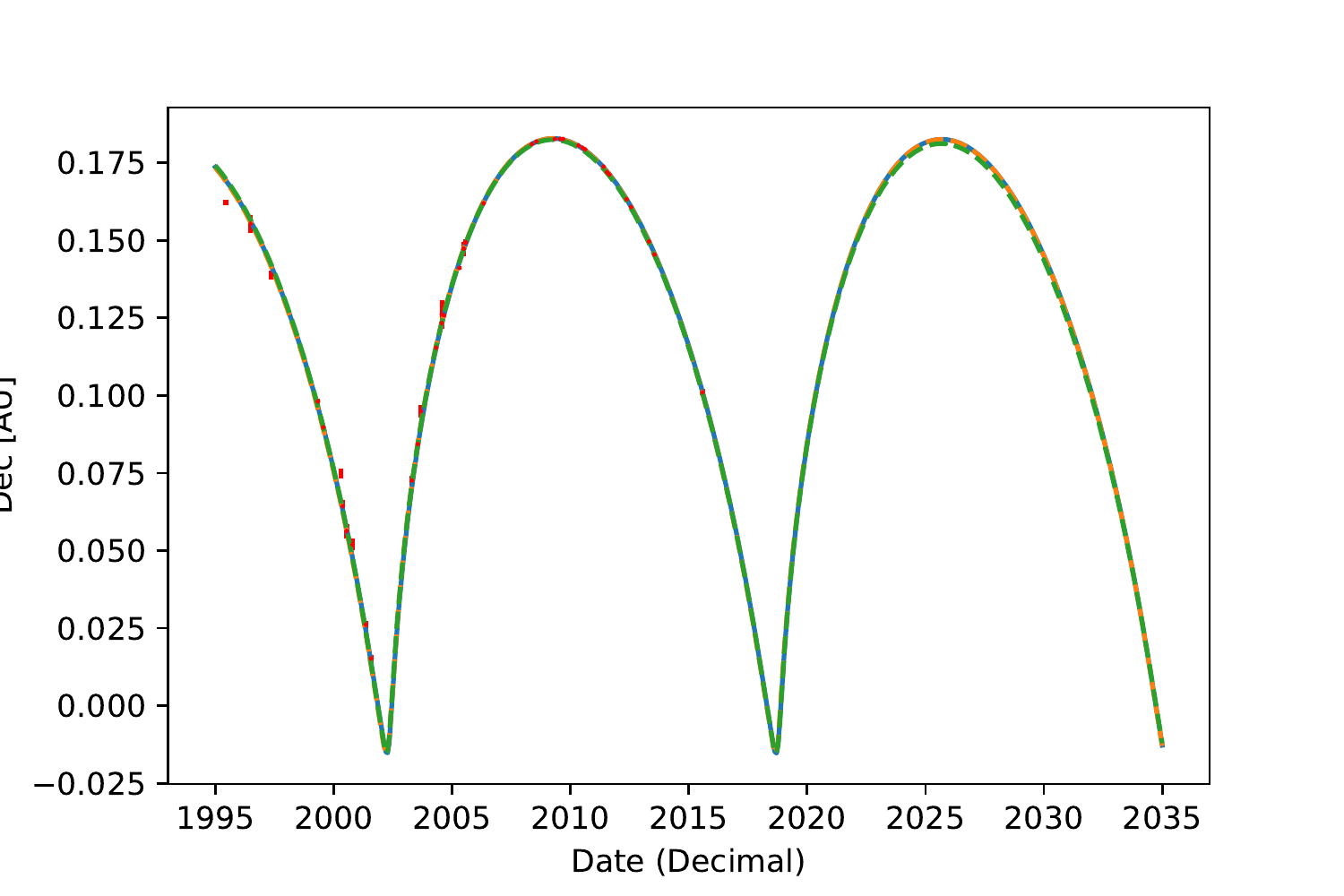}
\\
%\includegraphics[width=0.45\textwidth]{S2xfitRes.pdf}
%\includegraphics[width=0.45\textwidth]{S2yfitRes.pdf}

%\includegraphics[width=0.85\textwidth]{BestfitS2S38dec.pdf}
%\\
%\includegraphics[width=0.44\textwidth]{BestfitS2xy.pdf}
%\includegraphics[width=0.44\textwidth]{BestfitS38xy.pdf}

\caption{\it{The best fit of the S2 motion with the conformal and the disformal interactions.}} %The upper panel shows the fit with the observations and the lower panel shows the residuals.}}
 	\label{fig:numExamS2}
\end{figure*}
\begin{figure*}[t!]
 	\centering
\includegraphics[width=0.45\textwidth]{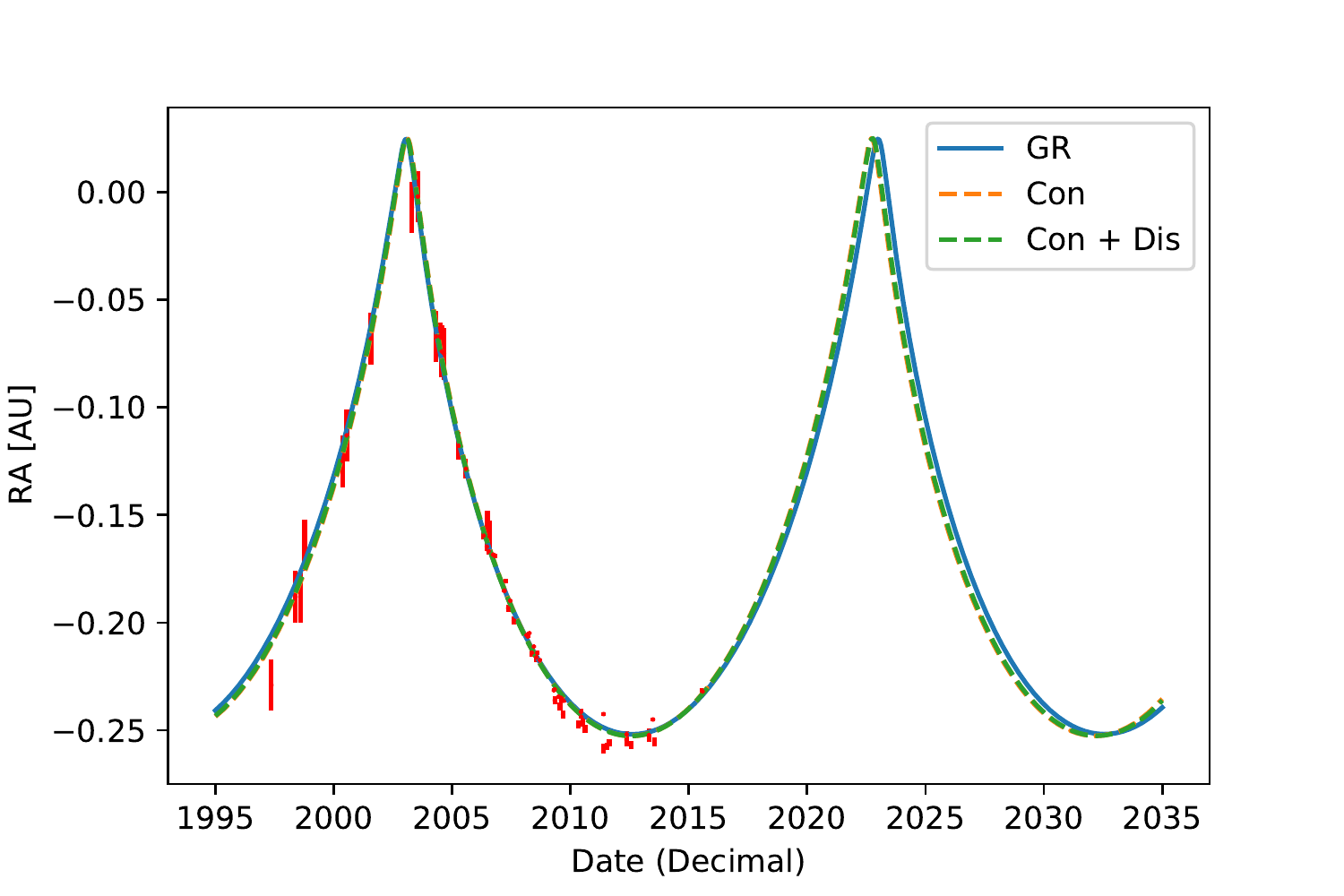}
\includegraphics[width=0.45\textwidth]{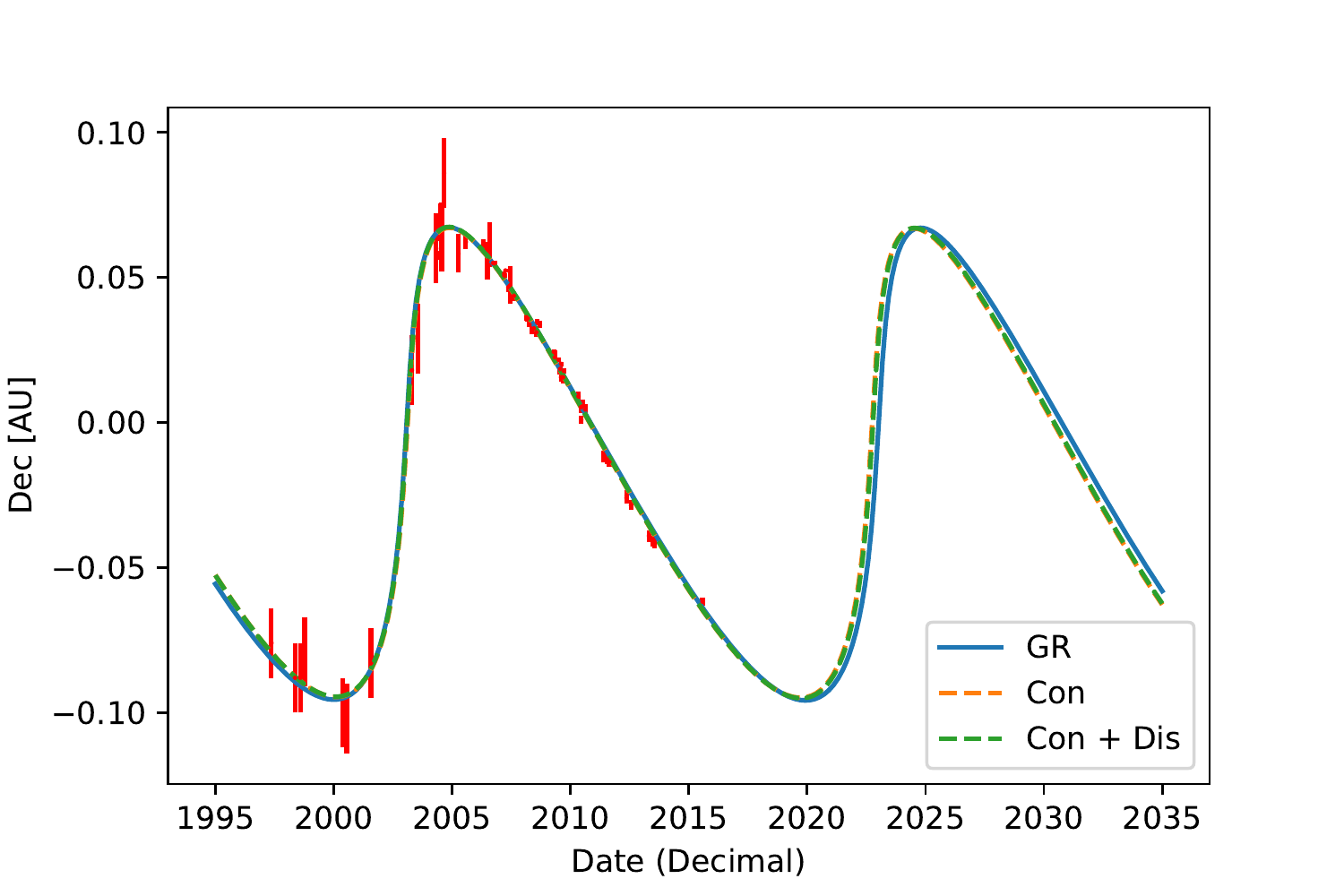}
\\

\caption{\it{The best fit of the S38 motion with the conformal and the disformal interactions.}} %The upper panel shows the fit with the observations and the lower panel shows the residuals.}}
 	\label{fig:numExamS38}
\end{figure*}

\begin{figure*}[t!]
 	\centering
%\includegraphics[width=0.85\textwidth]{BestfitS2S38dec.pdf}
%\\
\includegraphics[width=0.6\textwidth]{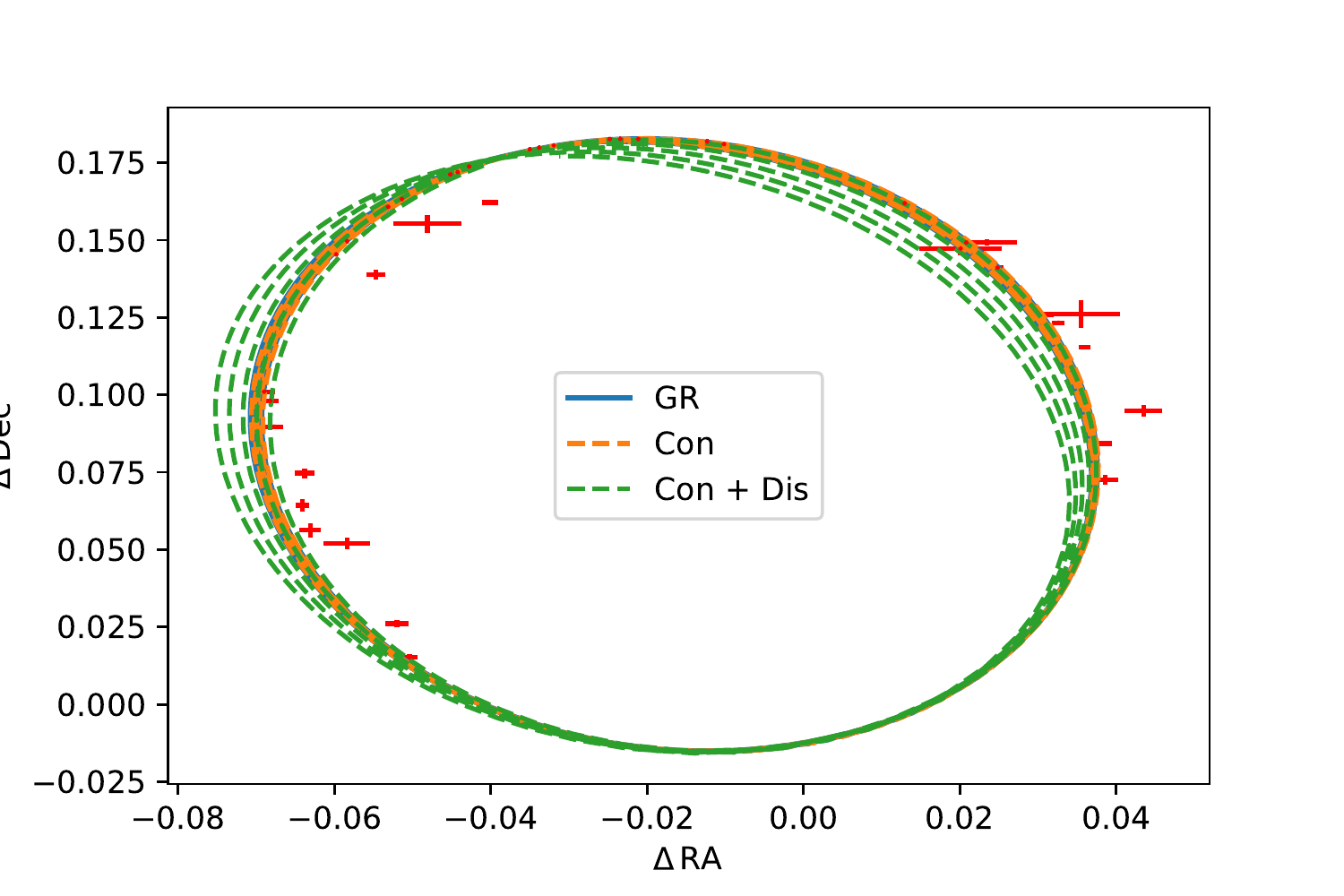}\\
\includegraphics[width=0.6\textwidth]{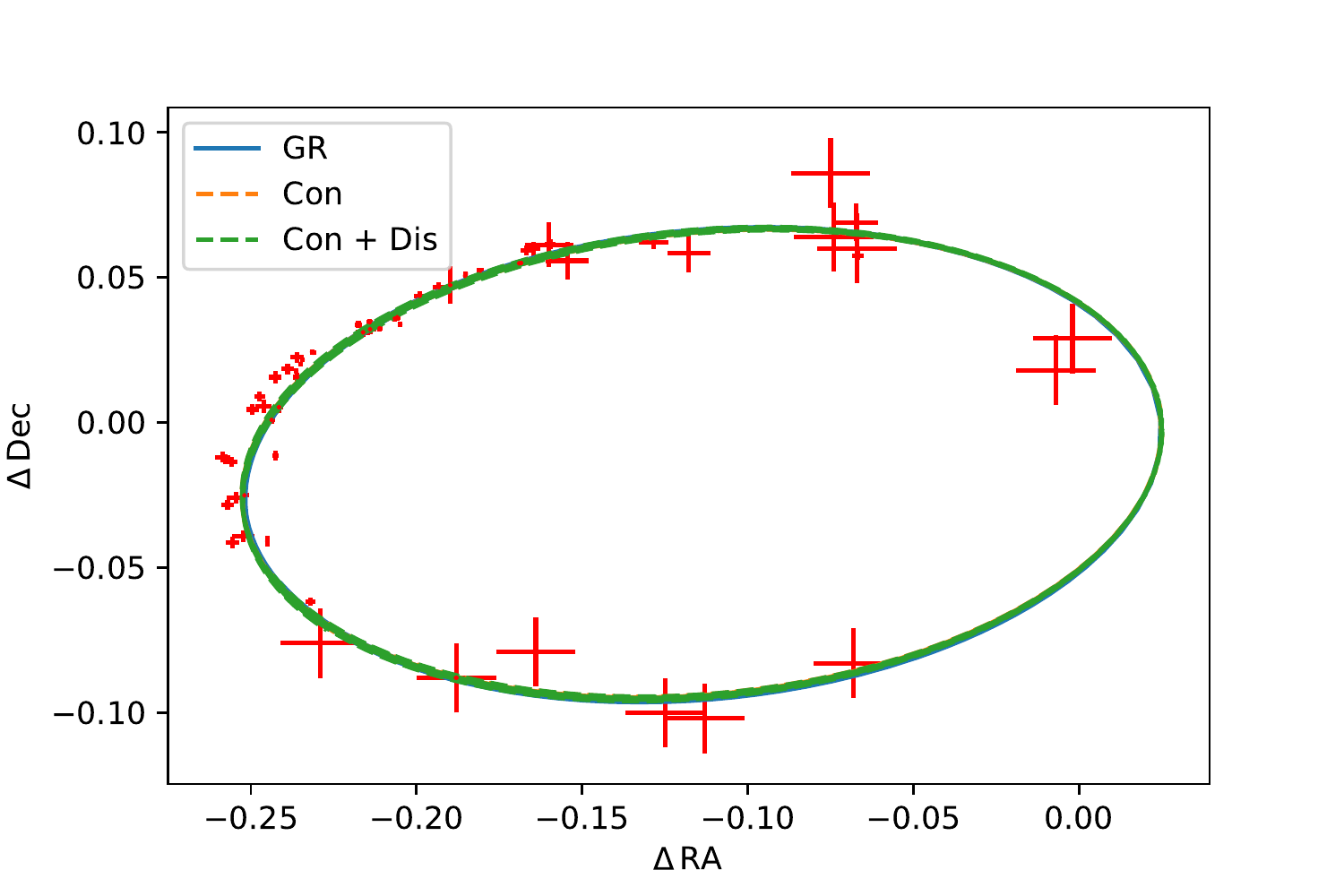}

\caption{\it{The best fit of the S2 (upper) and the S38 (lower) motions with the conformal and the disformal interactions. The evolution of the numerical simulation is for 60 years.}}
 	\label{fig:numExamS2}
\end{figure*}
\section{Methodology}
\label{sec:method}
We start the first iteration using a sampling of the initial 
position $(x_0, y_0)$ and velocity $(\dot{x}_0, \dot{y}_0)$ of the corresponding star in the orbital plane at the epoch at $1995$. The true positions $(x_i, y_i)$ and velocities $(\dot{x}_i, \dot{y}_i)$ at all successive observed epochs are then calculated by numerical integration of equations of motion (\ref{eq:eqMot}), and projected into the corresponding positions $(x_i^c, y_i^c)$ in the observed plane (apparent orbit). There are three angles that we take into account: $\Omega$ is longitude of the ascending node, $\omega$ is longitude of pericenter and $i$ is the inclination. The transformation from the reference frame to our frame is via the rotation matrix:
\begin{equation}
\begin{pmatrix}
l_1 & l_2\\
m_1 & m_2
\end{pmatrix}
\label{equ10}
\end{equation}
where the expressions for $l_1, l_2, m_1$ and $m_2$ depend on three orbital elements:
\begin{equation}
\begin{array}{l}
l_1=\cos\Omega\cos\omega-\sin\Omega\sin\omega\cos{i}, \\
l_2=-\cos\Omega\sin\omega-\sin\Omega\cos\omega\cos{i}, \\
m_1=\sin\Omega\cos\omega+\cos\Omega\sin\omega\cos{i}, \\
m_2=-\sin\Omega\sin\omega+\cos\Omega\cos\omega\cos{i}. \\
\end{array}
\label{equ11}
\nonumber
\end{equation}
We divided the location by the distance $d$ and rescaling the angle to $arcsec$ to be consistent with the observations obtained. We perform a Bayesian statistical analysis to constrain the four chameleon, or indeed general scalar-tensor theory, parameters, which we collectively denote by $\theta =  \{\vec{x}_0,\vec{\dot{x}}_0,M,d,\beta,\Lambda\,\omega \}$. The $\chi^2$ function entering the likelihood is given by
\begin{equation}
\chi^2(d,\Theta) = \sum_i \left(\frac{x_m^{(i)} (d) - x(\Theta)}{\Delta x_m^{(i)}}\right)^2 + \left(\frac{y_m^{(i)} (d) - y(\Theta)}{\Delta y_m^{(i)}}\right)^2.
\end{equation}
Our analysis uses publicly available astrometric and spectroscopic data that have been collected during the past thirty years. We use $145$ astrometric positions spanning a period from 1992.225 to 2016.38 from \cite{2017ApJ...837...30G}. The data come from speckle camera SHARP at the ESO New Technology Telescope \cite{1993Ap&SS.205....1H}, measurements were made using the Very Large Telescope (VLT) Adaptive Optics (AO).

\begin{figure*}[t!]
 	\centering
\includegraphics[width=0.8\textwidth]{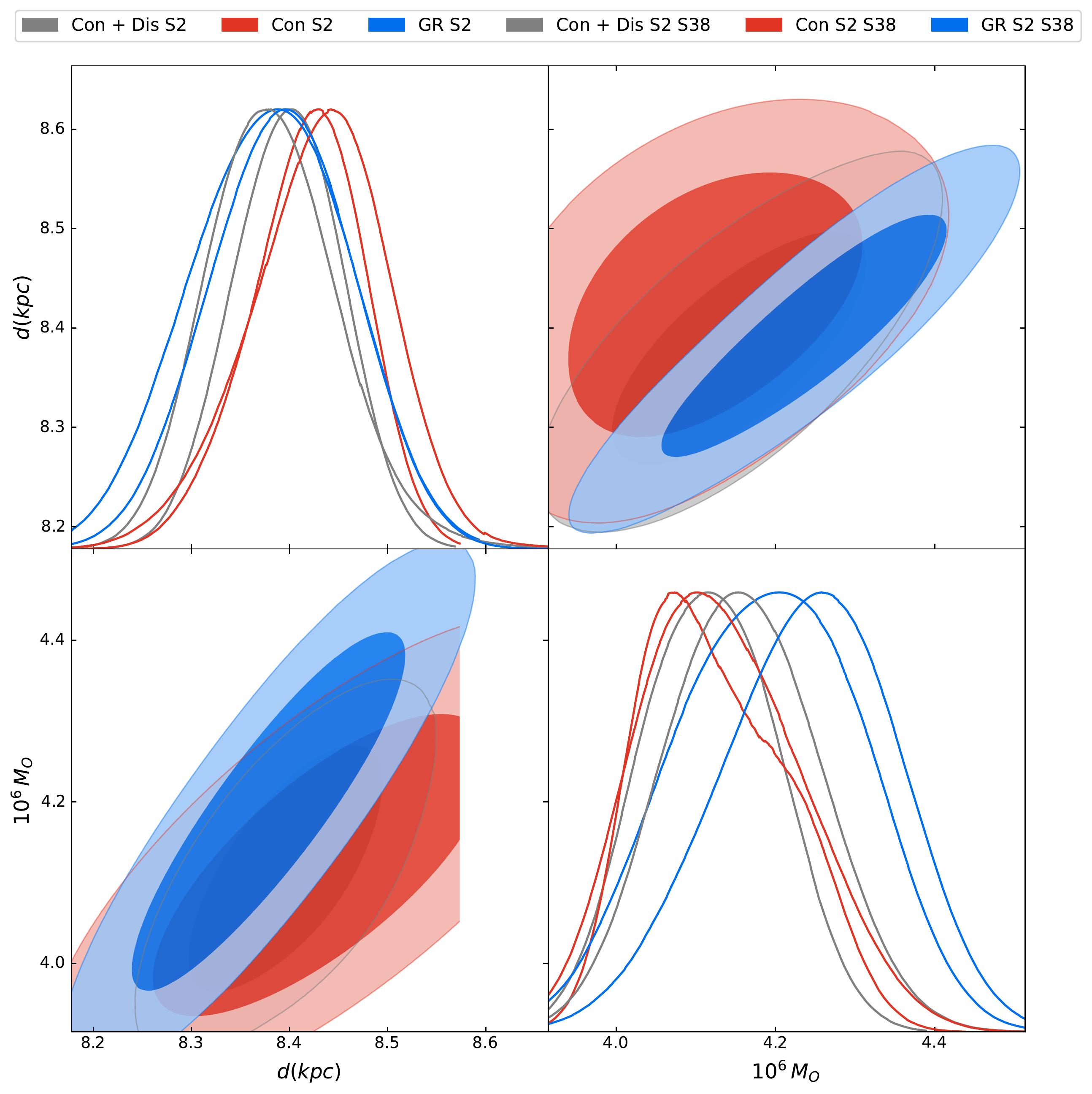}
\caption{\it{ {The posterior distribution for the S2 and the combined S2 and S38 motion for GR (blue), the conformal (red) and the disformal interactions (gray), with $1\sigma$ and $2\sigma$. Lower: the same combination for S2 and S38 combined analysis. } }} 
         
\end{figure*} 
\begin{table*}[t!]
\begin{tabular}{| c | c | c | c | c | c | c |}
\hline\hline
Parameter & GR S2 & Con S2 & Con + Dis S2 & GR S2 S38 & Con S2 S38 & Con + Dis S2 S38 \\
\hline\hline  
$M (10^6 M_\odot)$ & $4.18\pm 0.23 $ & $4.19\pm 0.16 $ & $4.25\pm 0.26$ & $4.13 \pm 0.08$ & $4.17\pm 0.10  $ & $4.18 \pm 0.1  $  \\
\hline
$d (kpc)$ & $8.38\pm 0.12 $ & $8.40\pm 0.13$ & $8.37 \pm 0.06$ & $8.42 \pm 0.07  $ &  $8.40 \pm 0.07  $ & $8.39 \pm 0.07  $\\
\hline
$\beta^2 (10^{-3})$ & - & $4.95 \pm 2.90 $ & $4.017 \pm 2.74$ & - & $5.97 \pm 2.78  $ & $5.4 \pm 3.0$ \\
\hline
$\Lambda^{-4} \left(10^{-2} \, AU/M_{\odot}\right)$ & - & - & $5.32 \pm 2.86$ & - & - & $5.21 \pm 2.90$ \\
\hline
$\Delta B_{ij}$ & - & -3.90 & 1.70 & - & -2.41 & -1.60  \\
\hline\hline  
\end{tabular} 
\caption{\it{The final results for the Mass of the Sgr A$\*$, the distance towards the black hole $d$ and the final values of the conformal $\beta$ and disformal $\Lambda^{-4}$ intestines. To complete our analysis we calculate the BE to compare the models. }}
\label{fig:1}
\end{table*}

Regarding the problem of likelihood maximization, we use an affine-invariant Markov Chain Monte Carlo sampler \cite{ForemanMackey:2012ig}, as it is implemented within the open-source packaged $\text{Polychord}$ \cite{Handley:2015fda} with the $\text{GetDist}$ package \cite{Lewis:2019xzd} to present the results. We test three different models: the GR solution, with conformal transformation and with the disformal transformation. We test these models with respect to the measurements of the S2 and the S38 stars. The S2 and S38 stars are the closest stars orbiting $Sgr A^{*}$ and the stars for which we have data.

 {In order to ensure that the conformal and the disformal coupling are subdominant, we choose the corresponding prior for our fit: $\beta^2  \in [0.;0.1]$  and $\Lambda^{-4}  \in [0.;0.03] \,\left(AU^3/M_{\odot}\right)$ for the conformal and the disformal strengths. Besides we use a unifrom prior on the distance towards the galactic center and the mass of $Sgr A^{*}$ as $d  \in [5.;10.] Kpc$ and $M \in [3.;6.] 10^{6} M_{\odot}$.} For simplicity we use the angles that reported by the Gravity collaboration as a Gaussian prior. For the rest we assume a uniform prior of the initial location and velocities.

\section{Results}
\label{sec:Res}
The posterior distribution of the S2 star fit is presented in the upper panel of \ref{fig:1} and the posterior distribution for the fit of the combined data of S2 and S38 is presented in the lower panel. 

The obtained mass of the $Sgr A^{*}$ is changed a bit with the S38 measurements: from $4.18\pm 0.23 \, 10^6 M_\odot$ for GR, $4.19\pm 0.16 \, 10^6 M_\odot$ for the confromal interaction and $4.25\pm 0.26 \, 10^6 M_\odot$ with the disformal interaction, into  $4.13 \pm 0.08 \, 10^6 M_\odot$ for GR,  $4.13 \pm 0.08 \, 10^6 M_\odot  $  for the conformal interaction $4.18 \pm 0.1 \,10^6 M_\odot $ with the disformal interaction. The distance towards $Sgr A^{*}$ is a bit decreases: from $8.38\pm 0.12 kpc$ for GR, $8.40\pm 0.13 kpc$ for the conformal interaction and $8.37 \pm 0.06 kpc$ with the disformal interaction, into  $8.42 \pm 0.07  kpc$ for GR, $8.40 \pm 0.07 kpc$  for the conformal interaction $8.39 \pm 0.07  kpc$.

The constraint on the conformal interaction parameter $\beta^2$ is around $10^{-3}$: $4.95 \pm 2.90 \, \cdot 10^{-3}$ for the S2 star and $4.017 \pm 2.74 \, \cdot 10^{-3} $ for the combined with S38 star. Including the disformal interaction yields $5.97 \pm 2.78 \, \cdot 10^{-3} $ for the S2 star and $5.4 \pm 3.0 \, \cdot 10^{-3}$ for the combined stars. The case of $\beta^2$ goes to zero is the GR case, however this value is outside of the  $1\sigma$ mean value. This gives potential support for that this additional interaction to exist. The constraint on the disformal interaction parameter $\Lambda^{-4}$ is around $< 10^{-3}$: $5.32 \pm 2.86 y^{-2}$ for the S2 star and $5.21 \pm 2.90 y^{-2}$ for S2 combined with S38 star. The case of $\Lambda$ goes to infinity is the GR case. %From the current values $\Lambda$ is around $10^{20}$ and more, giving a very strong constraint on this parameter.

Finally, we can further quantify the relative ability of the conformal and the disformal interactions model to describe the various data sets {\it w.r.t.} the GR cosmology using the Bayes ratio.  Given a data set $\mathcal{D}$, the probability of a certain model $M_i$ to be the best one among a given set of models $\{M\}$ reads, 
\begin{equation}\label{eq:BayesTheorem}
P(M_i|\mathcal{D})=\frac{P(M_i)\mathcal{E}(\mathcal{D}|M_i)}{P(\mathcal{D})}\,,
\end{equation}
where $P(M_i)$ is the prior probability of the model $M_i$ and  $P(\mathcal{D})$ the probability of having the data set $\mathcal{D}$, with the normalization condition $\sum_{j\in\{M\}}P(M_j)=1$. The quantity $\mathcal{E}(\mathcal{D}|M_i)$ is the so-called marginal likelihood or evidence. If the model $M_i$ has $n$ parameters $p^{M_i}_n$, the evidence reads, 
\begin{equation}\label{eq:evidence}
\mathcal{E}(\mathcal{D}|M_i)=\int \mathcal{L}(\mathcal{D}|\vec{p}^{M_i},M_i)\pi(\vec{p}^{M_i}) d^np^{M_i}\,,
\end{equation}
with $\mathcal{L}(\mathcal{D}|\vec{p}^{M_i},M_i)$ being the likelihood and $\pi(\vec{p}^{M_i})$ the prior of the parameters entering the model $M_i$. If we compare the models by assuming equal prior probability for both of them, then we find that the ratio of their associated probabilities is directly given by the ratio of their corresponding evidences, i.e.
\begin{equation}\label{eq:BayesRatio}
\frac{\mathcal{E}(\mathcal{D}|{\rm con+ dis})}{\mathcal{E}(\mathcal{D}|GR)} \equiv B_{ij}\,.
\end{equation}
This is known as Bayes ratio and is the quantity we are interested in \cite{doi:10.1080/01621459.1995.10476572}. In order to complete the analysis we compare the BE of the models we use the values from the Polychord. The BE values are summarized in fig \ref{fig:1}. In the first case we take only the S2 measurements. The difference between the BE of the GR case and the conformal model is $-3.90$ and the difference between the BE of the GR case and the conformal and the disformal model is $1.70$. Therefore the is a slight preference for GR vs. the conformal coupling, and an indistinguishable preference for conformal with the disformal coupling vs. GR. In the second case that we test S2 with the S38 stars The difference between the BE of the GR case and the conformal model is $-2.41$ and the difference between the BE of the GR case and the conformal and the disformal model is $-1.60 $.  Combining the observations of the S38 star yields a slight preference for GR.

\section{Combined Constraints}
\label{sec:ComCont}

 {In order to complete our discussion on the interactions in two body motion, we add the solar system constraint \cite{Ip:2015qsa}. \cite{Brax:2018bow} gives the precession term for the conformal and the disformal contribution as:}
\begin{equation}
\Delta\theta = \frac{2 \pi G_N M }{a c^2 (1-e^2)}\left(3 - 2 \beta^2  + \frac{5 \beta^2 M c^2}{2 \pi \Lambda^4 a^3 (1-e^2)^3 } \right).
\end{equation}
 {Correspondingly, the $\chi^2$ from some object reads:}
\begin{equation}
\chi^2 = \left(\frac{\theta_{ob} - \theta\left(\beta, \Lambda\right)}{\sigma_{\theta}}\right)^2.
\end{equation}
 {We include the precession data of the solar system from \footnote{https://nssdc.gsfc.nasa.gov/planetary/factsheet/}.
The combined constraint is approached by using the combined likelihood:
\begin{equation}
\chi^2 = \chi^2_{S2} + \chi^2_{S38} + \chi^2_{\theta,SolSys}, 
\end{equation}
where $\chi^2_{S2}$ and $\chi^2_{S38}$ include the measurements from the galactic star centre and $\chi^2_{\theta,SolSys}$ include the precession measurements from the solar system.}

 {The last bound that we take into account is the Cassini bound \cite{Bertotti:2003rm}. Cassini gives a bound of $\beta^2 \leq 4 \cdot 10^{-5}$ for the $\beta$ parameter without any special bound on $\Lambda$. In order to include the Cassini bound, we use a smaller prior on $\beta$. }

 {Fig. (7) shows the combined analysis for the conformal and the disformal couplings for different cases. The table below summarizes the final values. We see that adding the solar system constraints, without Cassini, gives very little changer to the modified gravity parameters. However, adding the Cassini bound reduces the final value of $\beta^2$ into $\sim 10^{-5}$ and the value of $\Lambda^{-4}$ into $\sim 5\cdot 10^{-2} \, AU^3/M_{\odot}$ which is $\Lambda \sim 0.08 MeV$. }
%\textcolor{blue}{What's the overall constraint on $\Lambda$ in MeV? }

\begin{figure*}[t!]

 	\centering

\label{fig:final}

\includegraphics[width=0.8\textwidth]{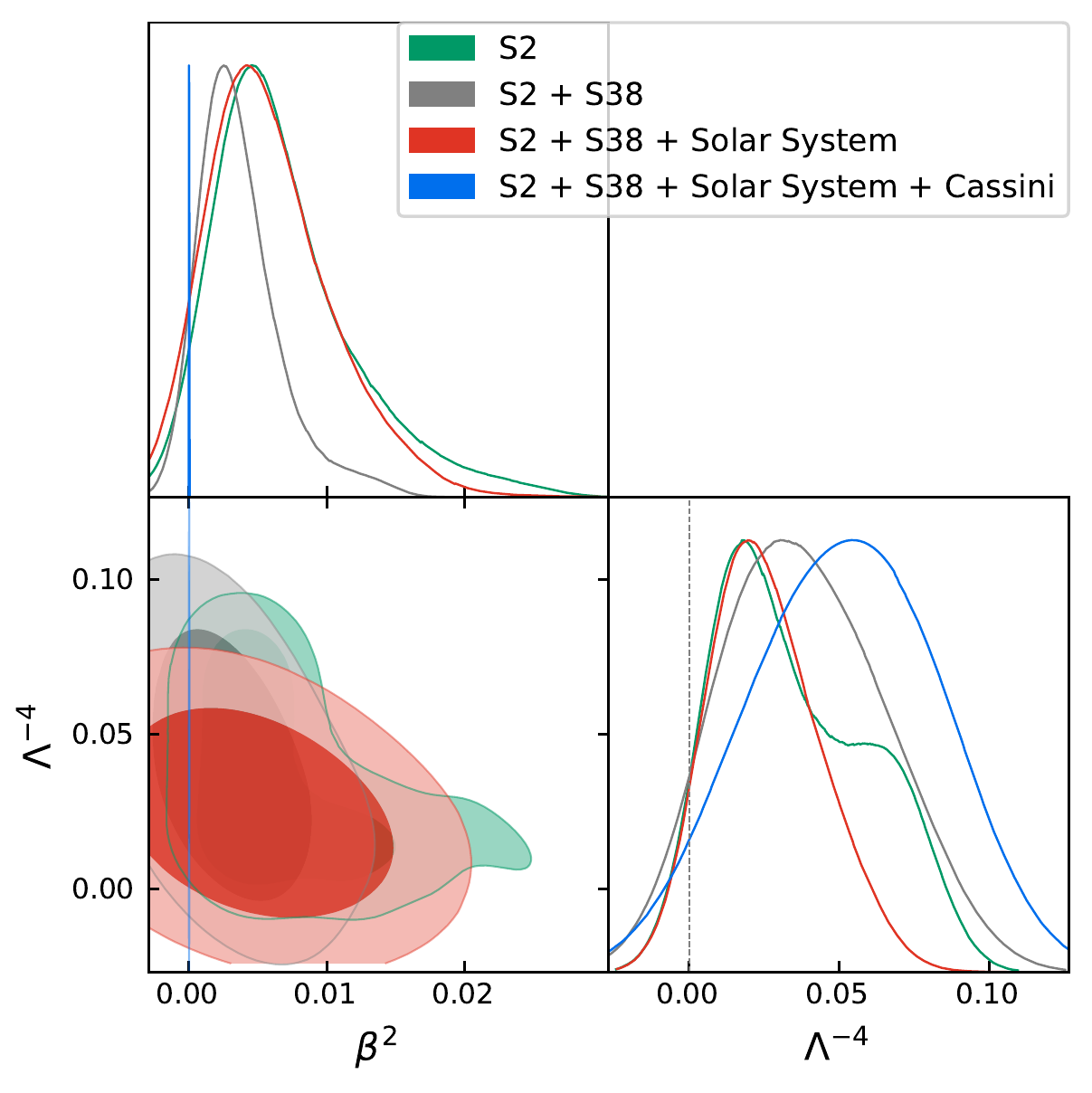}

\begin{tabular}{| c | c | c |}
\hline\hline
Parameter &  S2 + S38 + Solar System & S2 + S38 + Solar System + Cassini \\
\hline\hline  
$\beta^2$ & $\left(5.904 \pm 4.71\right)\cdot 10^{-3}$ & $\left(2.080 \pm 1.186\right)\cdot 10^{-5}$ \\
\hline
$\Lambda^{-4} \left(10^{-2}\,AU^3/M_{\odot}\right)$ & $2.571 \pm 1.599  $ & $5.313\pm2.604$ \\
\hline\hline  
\end{tabular} 

\caption{\it{ The contour plot for the conformal and the disformal couplings for the S2, S38 stars, the solar system and the Cassini bound. The point $(0,0)$ corresponds to GR with no additional interaction. The combined data set gives stronger constraint on the values of the extended parameters.}}

\end{figure*}

\section{Discussion}
\label{sec:Dis}
We have investigated modified gravity theories in the strong gravity regime near the galactic centre $Sgr A^{*}$ using the observations of the S2 and S38 stars. To our knowledge this is the first time such theories have been investigated in the strong gravity regime. The modified gravity scalar field is both conformally and disformally coupled to matter and the observations allow both types of interaction to be constrained.  Our formalism is very general using a PN expansion. As noted previously, the disformal coupling has an effect only when it combines with a conformal interaction. In this case, the disformal coupling leads to a change in the advance of perihelion for a light body and modifies the effective metric which governs the evolution of two interacting bodies. The orbital measurements around the galactic star centre $Sgr A^{*}$, in particular the locations and the velocities of the S2 and the S38 stars, is used in our constraints. Using a MCMC simulation yields a bound on the parameters, $\beta$ and $\Lambda$ of the modified gravity theory. In our analysis we note that the BE gives a slight preference for GR.

 {The bound on the conformal and disformal couplings from different systems is widely constrained. Such a range for $\Lambda$ is constrained from the Eotwash experiment with the lower bound of $\Lambda > 0.07 MeV$ and $\beta^2 < 4 \cdot 10^{-5}$ from the Cassini experiment \cite{Bertotti:2003rm}. The posterior distribution for $\beta^2$ is about $10^{-5}$ and the bound of $\Lambda$ yields a range of $ \sim 0.08 MeV$. The bounds we have obtained from the galactic star centre with the Cassini experiments gives a strong statement on the interactions of DE in the strong gravity regime of the galactic centre.}

 {In future it would be interesting to add other cosmological measurements such from those arising from the expansion history of the Universe and structure formation data. The cosmological data require more specific form of potentials and conformal and disformal couplings in order to make a complete combined analysis. Such analysis is left for future work. }

\acknowledgments
We thank Wyn Evans, Eduardo I. Guendelman, Salvatore Cappoziello and Philippe Brax for discussions and insightful comments. D.B gratefully acknowledge the support from the Blavatnik and the Rothschild fellowships. We have received partial support from European COST actions CA15117 and CA18108 and STFC consolidated grants ST/P0006811 and ST/T0006941.

\bibliographystyle{apsrev4-1}
\bibliography{ref}

\end{document}